\documentclass[a4paper,10pt,twocolumn,aps,prl,superscriptaddress,showpacs]{revtex4}
\usepackage{graphicx}

\begin{document}

\title{Hydrodynamics and the fluctuation theorem}
\author{M.~Belushkin}\email{maxim.belushkin@epfl.ch}
\affiliation{Institute of Theoretical Physics, Ecole Polytechnique F\'{e}d\'{e}rale de Lausanne (EPFL), CH-1015 Lausanne}
\author{R.~Livi}\altaffiliation[on leave of absence from ]{Dipartimento di Fisica, Universita' di Firenze, I-50019 Sesto Fiorentino}
\affiliation{AGM (CNRS UMR 8088) and LPTM (CNRS UMR 8089), Universit{\'e} de Cergy-Pontoise, F-95302 Cergy-Pontoise}
\author{G.~Foffi}\email{giuseppe.foffi@epfl.ch}
\affiliation{Institute of Theoretical Physics, Ecole Polytechnique F\'{e}d\'{e}rale de Lausanne (EPFL), CH-1015 Lausanne}

\begin{abstract}
The fluctuation theorem is a pivotal result of statistical physics. It quantifies the probability of observing
fluctuations which are in violation of the second law of thermodynamics. More specifically,
it quantifies the ratio of the probabilities of observing entropy-producing and entropy-consuming fluctuations
measured over a finite volume and time span in terms of the rate of entropy production in the system, the
measurement volume and time. We study the fluctuation theorem in computer simulations of planar shear flow.
The simulations are performed employing the method of multiparticle collision dynamics which captures both thermal fluctuations and hydrodynamic interactions. The main outcome of our analysis is that the fluctuation theorem is verified at any averaging time provided that the measurement volume exhibits a specific dependence on a hydrodynamic time scale.
\end{abstract}

\pacs{47.52.+j, 47.11.-j, 05.70.Ln}
\maketitle
The fluctuation theorem states that, in a system in steady-state conditions, the probability of observing some amount of entropy consumption relative to the probability of observing the same amount of entropy production is exponentially suppressed. More specifically,~\cite{Evans:1993,Gallavotti:1995,Evans:2002}
\begin{equation}
 \frac{P(\overline{\Omega}_{\tau} = -R)}{P(\overline{\Omega}_{\tau} = +R)} = \mathrm{e}^{-R \tau} \, ,
\label{eq:fluctuation_theorem}
\end{equation}
where $R$ is the average value taken by the entropy production rate $\Omega$ over the averaging time $\tau$,
$\overline{\Omega}_{\tau} = R = \frac{1}{\tau} \int\limits_{0}^{\tau} \Omega(t)\mathrm{d}t$.
\par
Among numerous attempts to verify the validity of the fluctuation theorem both directly~(\ref{eq:fluctuation_theorem}) and in one of its integral forms~\cite{Lepri:1998,Lepri:2000,Wang:2002,Carberry:2004,Schuler:2005,Douarche:2006,Ciliberto:2010}, 
it has been observed that the fluctuation theorem in its direct form~(\ref{eq:fluctuation_theorem}) does not appear to hold in the presence of hydrodynamic interactions~\cite{Bonetto:2001}, although, of course, it is still satisfied in an integral form.
\par
In this contribution, we demonstrate the validity of the fluctuation theorem in the presence of hydrodynamic interactions, and
show that the origin of the previously observed apparent inconsistency has a well-defined physical interpretation which lies in correlations and hydrodynamic transport within the fluid. More specifically, we show that in the presence of hydrodynamics, the entropy production rate is a global effect because of correlations. When these correlations are eliminated, the extensivity of entropy production is lost. Our results not only serve as a test of the fundamental laws of thermodynamics, but also provide crucial insight for future experiments aimed at putting the fluctuation theorem to the test.
\par
\begin{table}[t!]
 \begin{tabular}{|c|c|c|c|c|c|}
  \hline
  Setup & $L [a]$ & $\rho [\frac{m}{a^{3}}]$ & $\delta t [\sqrt{\frac{ma^{2}}{k_{B}T}}]$ & $\eta [\frac{k_{B}T m}{a^{4}}]$ & $\gamma [\gamma_{0}]$\\ \hline
  1 & 8 & 5 & 0.05 & 7.47 & 1\\ \hline
  2 & 16 & 5 & 0.05 & 7.47 & 1\\ \hline
  3 & 16 & 5 & 0.1 &  3.96 & 1\\ \hline
  4 & 16 & 10 & 0.05 & 16.67 & 1\\ \hline
  5 & 16 & 10 & 0.1 &  8.7 & 1\\ \hline
  6 & 8 & 10 & 0.05 & 16.67 & 5 \\ \hline
  $7^{\dagger}$ (no HI) & 8 & 5 & 0.1 & - & 1\\ \hline
 \end{tabular}
 \caption{\label{table:parameters}List of setups for which simulations have been performed together with the
respective simulation parameters. Setup $7^{\dagger}$ is constructed such that hydrodynamic interactions (HI)
are not present. The viscosities ($\eta$) are calculated using the analytical formulas of Ref.~\cite{Gompper:2009}.
For setup  $7^{\dagger}$ the viscosity cannot be evaluated in this way due to the implementation of a non-hydrodynamic system.
The shear rates are given in units of $\gamma_{0}=0.00625 \sqrt{k_{B}T/ma^{2}}$.}
\end{table}
Following Ref.~\cite{Bonetto:2001}, we consider steady-state planar shear flow.
We choose shear to be in the $x$ direction with a shear rate
$\gamma=\mathrm{d}v_{x} / \mathrm{d}y$.
Since in this case the entropy production rate can be expressed in terms of the stress tensor
$\sigma_{xy}$, $\Omega = \sigma_{xy}\gamma V/k_{B}T$, where $V$ is a volume and $k_{B}T$ denotes the thermal energy,
the fluctuation theorem can be written as~\cite{Evans:2002}
\begin{equation}
 {\cal P}(\tau,V) = - \tau V\, ,
\label{eq:shear_fluctuation_theorem}
\end{equation}
where we have defined the probability function
\begin{equation}
{\cal P}(\tau,V) = \frac{k_{B}T}{A \gamma} \ln \frac{P(\overline{\sigma}_{xy,\tau}=-A)}{P(\overline{\sigma}_{xy,\tau}=+A)}\, .
\label{eq:PoftauV}
\end{equation}
\par
A key statement of the fluctuation theorem is that ${\cal P}(\tau,V)$ is independent
of the specific properties of the system, i.e. the viscosity or the shear rate. Rather, ${\cal P}(\tau,V)$ depends on
an extensive variable $V$ and a measuring time $\tau$. The extensive variable $V$ is related to the volume in which measurement of
entropy production is performed.
\par
To study the fluctuation theorem~(\ref{eq:shear_fluctuation_theorem}), we perform computer simulations employing the method
of multiparticle collision dynamics (MPC). This method is known to correctly capture both thermal fluctuations and hydrodynamic
interactions on coarse-grained length- and time-scales~\cite{Padding:2006,Gompper:2009}.
It has been successfully applied to a wide range of steady-state problems, i.e. colloids in shear flow~\cite{Hecht:2006}, polymers in shear flow~\cite{Nikoubashman:2010}, vesicles in shear flow~\cite{Noguchi:2007} and colloidal rods in shear flow~\cite{Ripoll:2008}.  In MPC, the solvent is modeled as a set of point particles with continuous coordinates and velocities. The system evolves in discrete time steps. At each step, the solvent particles undergo propagation and collision. During propagation, the coordinates $\vec{r}_{i}$ of each particle are updated according to its velocity $\vec{v}_{i}$ and a time interval $\delta t$,
$\vec{r}_{i} \rightarrow \vec{r}_{i} + \vec{v}_{i}\delta t$. During the collision step, all particles are sorted into
collision cells, and within each cell their velocities are rotated with respect to the center of mass velocity
by an angle $\alpha$ around a random axis. To guarantee Galilean invariance, the structure of the collision grid is shifted randomly before each collision step. The time interval $\delta t$ denotes the time between successive collisions within the solvent and is therefore directly related to the transport coefficients of the solvent. Thus, we denote $\delta t$ as the collision time.
\par
The general framework of calculations is as follows. Given a steady-state evaluation of the viscous stress
tensor $\sigma_{xy}(t)$, where each measurement of the time series is performed within a measurement volume
$V$ and over a measurement time $\tau$, we construct the probability distribution $P(\sigma_{xy})$ of all the
values in the time series. We then average the probability distribution over 100 simulation runs with different
initial conditions. It is then possible to test the fluctuation theorem~(\ref{eq:shear_fluctuation_theorem}).
\par
We fix the simulation parameters to
values typical of collective, fluid-like behavior~\cite{Ripoll:2005}. As normalization, we choose the
solvent particle mass $m=1$, the collision cell size $a=1$ and the thermal energy $k_{B}T=1$. Mass is measured
in units of $m$, length in units of $a$ and time in units of $t_{0}=\sqrt{m a^{2} / k_{B}T}$. The simulation
box sizes $L=8a$ and $L=16a$ are intentionally chosen small to enhance fluctuations. The collision angle $\alpha$ is
taken to be $130^{\mathrm o}$. The remaining parameters are summarized in Table~\ref{table:parameters} for
the several setups for which simulations have been performed.
\par
We employ Lees-Edwards boundary conditions to generate planar shear flow~\cite{Lees:1972}. In order to achieve
steady-state conditions, the system is kept at constant temperature by a
stochastic cell-level thermostat~\cite{Huang:2010}. All simulations run for
5000 steps before measurements begin to be made to guarantee steady-state conditions.
\par
\begin{figure}[t!]
 \begin{tabular}{cc}
  \includegraphics[width=0.45\linewidth]{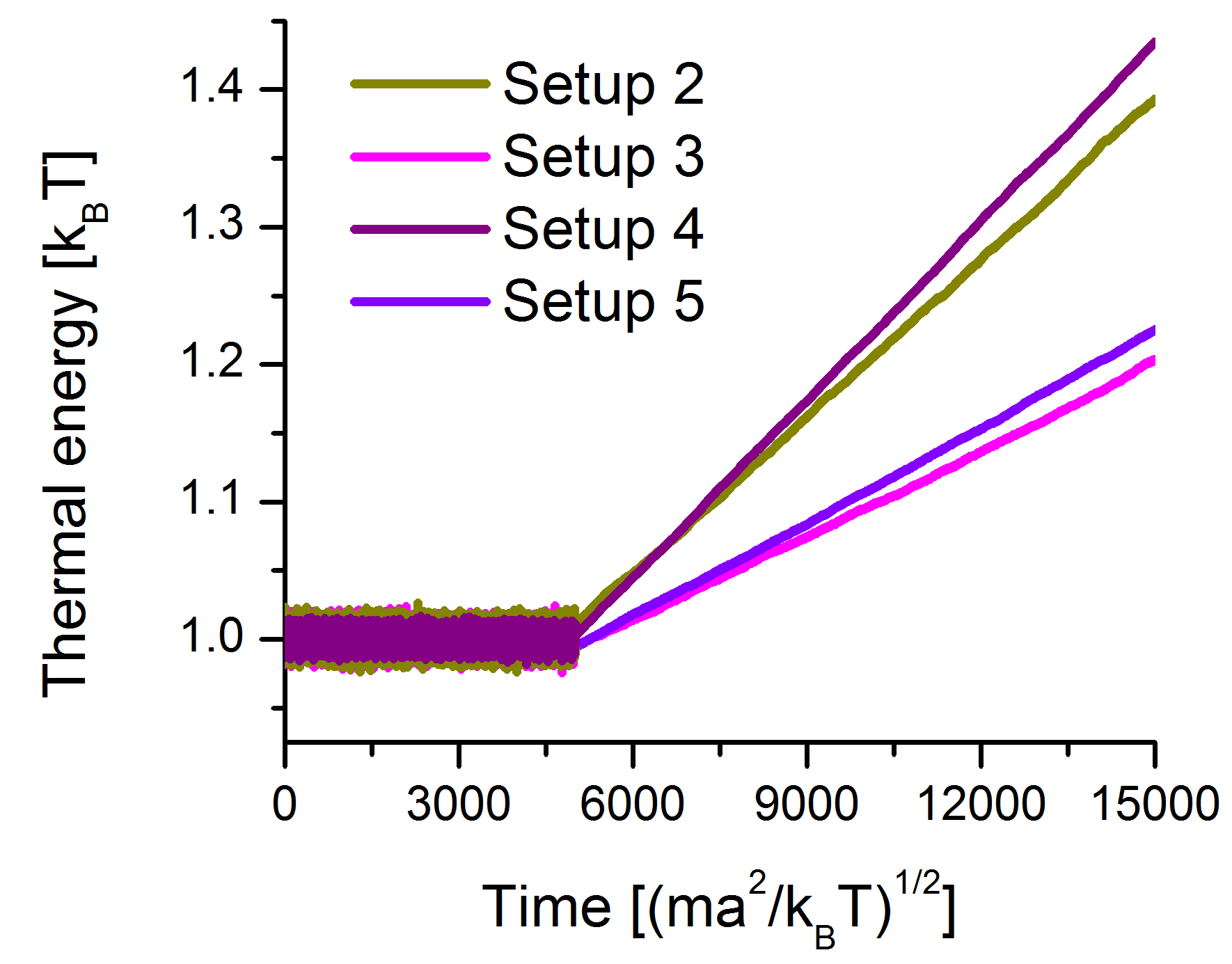}
  \includegraphics[width=0.45\linewidth]{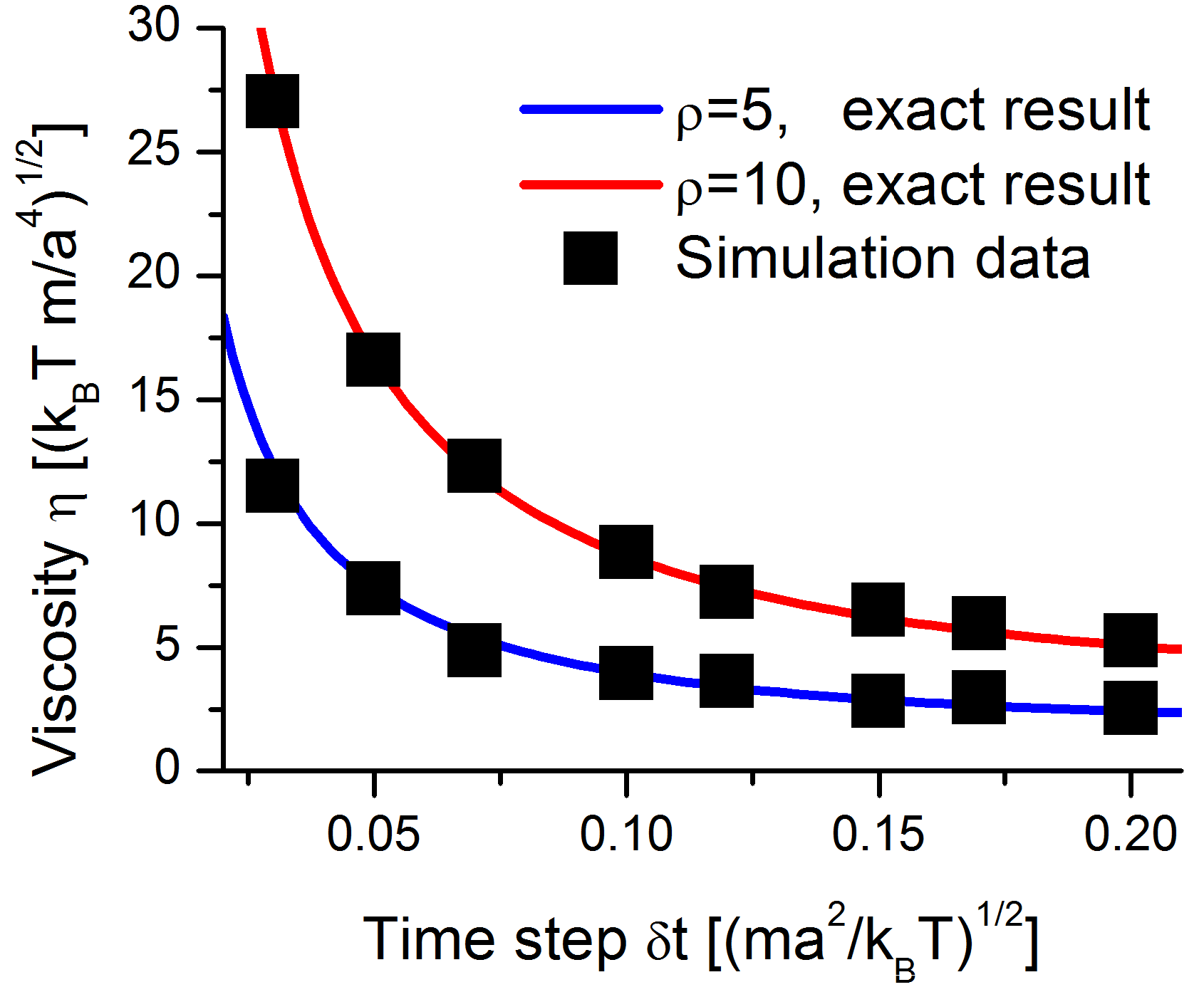}
 \end{tabular}
  \caption{\label{fig:viscosity_measurement}(Color online) Left: thermal energy measured in setups 2-5 with ($t<5000$) and without ($t>5000$) a thermostat.
   Right: viscosity measured through the rate of temperature increase of the fluid (boxes) compared to the exact analytical results~\cite{Gompper:2009} (solid lines).}
\end{figure}

As the first step of the present work, we test an integral form of the fluctuation theorem. To this end, we employ a simple method to relate the viscosity of the fluid, $\eta$, to the rate of temperature increase in the system undergoing shear flow in the absence of a thermostat~\cite{Naitoh:1979},
\begin{equation}
C_{V}\frac{dT}{dt} = \eta \gamma^{2}
\end{equation}
where $C_{V}$ denotes the volumetric heat capacity.
We observe that in the absence of a thermostat, the temperature of the system increases linearly (Fig.~\ref{fig:viscosity_measurement}, left). The results for the viscosity (Fig.~\ref{fig:viscosity_measurement}, right)
are in excellent agreement with the exact analytical result~\cite{Gompper:2009} which, in our case, gives the shear viscosity as a function of the solvent density $\rho$ and the collision time $\delta t$. It is well-known that viscosity determined from the long-time limit of the viscous stress tensor is in excellent agreement with the theory~\cite{Kikuchi:2003}. Here, we have demonstrated that the relation between entropy production and a transport coefficient holds, confirming that overall entropy production is captured quantitatively~\cite{Gallavotti:1996}. This result is not {\it a priori} evident despite the existence of an H-theorem for MPC~\cite{Malevanets:1999}. We remark that the novel method for the determination of viscosity using the rate of temperature increase of the MPC fluid undergoing shear flow presented here requires, in principle, only very short simulation runs, as opposed to the determination of the viscosity from the long-time limit of the viscous stress tensor~\cite{Kikuchi:2003}.
\par
\begin{figure}[t!]
\begin{center}
  \includegraphics[width=.9\linewidth]{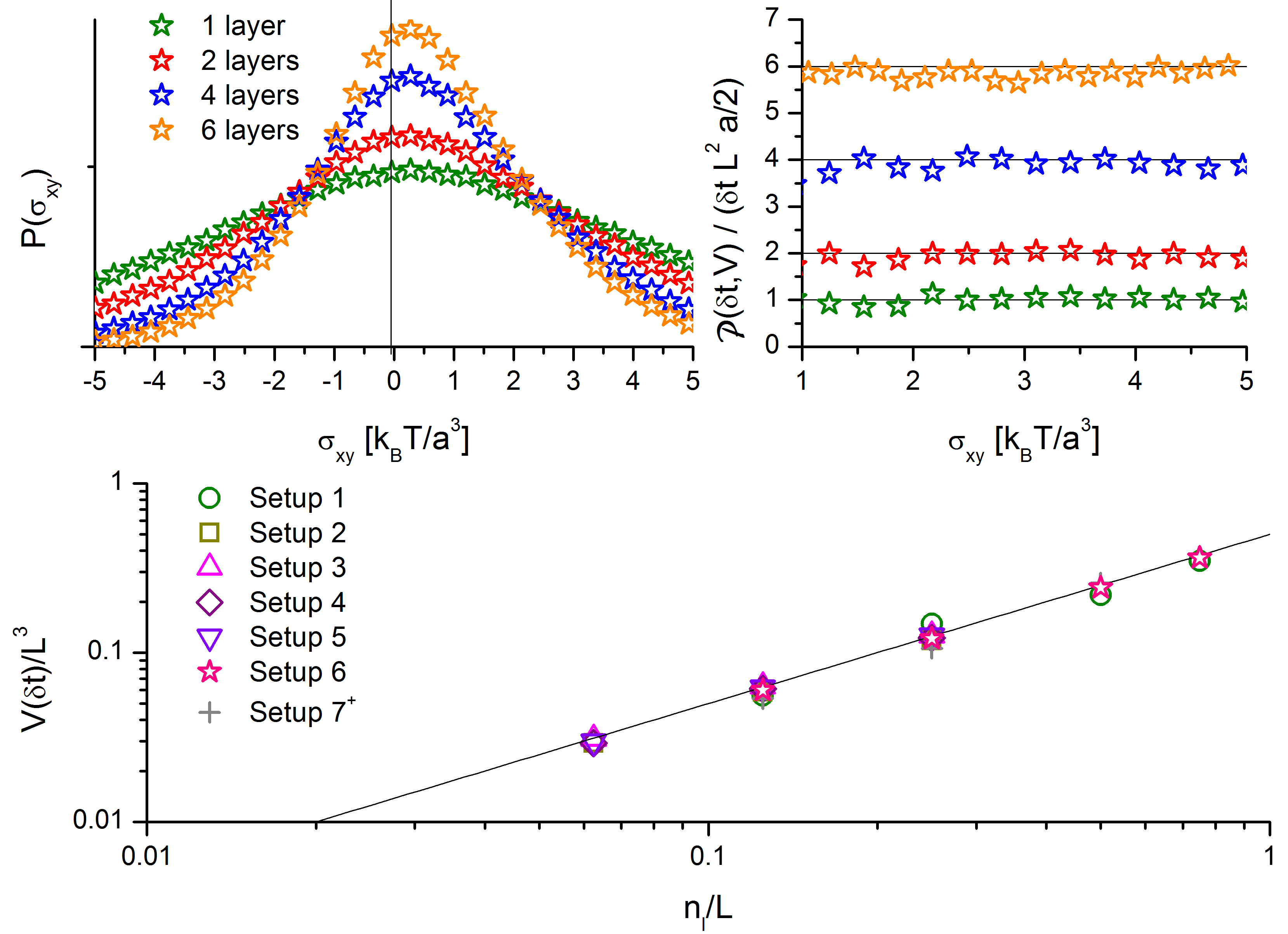}
\end{center}
\caption{\label{fig:scaling_volume}(Color online) Top left: measured probability distributions $P(\sigma_{xy})$ for setup 6 for one (green), two (red), four (blue) and six (orange)
layers over the shortest averaging time $\tau=\delta t$ are not zero-centered and not symmetric. They become increasingly peaked for increasing measurement volumes, as expected.
Top right: the corresponding probability functions ${\cal P}(\tau,V)$ normalized by the measurement time $\tau=\delta t$ and the volume of a single layer $L^{2}a/2$ are constant and correctly reflect the number of measurement layers.
Bottom: the resulting scaling of the observation volume $V(\tau)$ defined in Eq.~(\ref{eq:effective_volume}) with the number of measurement planes $n_{l}$ is as predicted by the fluctuation theorem for all the simulation setups. Different symbols denote different setups.}
\end{figure}
Having verified the integral form of the fluctuation theorem~\footnote{Furthermore, MB and GF have verified that the spectral properties of the thermal fluctuations within MPC are consistent with theory. The results will be communicated in a separate publication.}, we proceed to test its direct form~(\ref{eq:shear_fluctuation_theorem}).
To this end, we measure entropy production by measuring the viscous stress tensor $\sigma_{xy}$
as the $x$-component of momentum crossing a plane of constant $y$ per unit area and per unit time at each
simulation step~\cite{Kikuchi:2003}. We choose the measurement planes in such a way that on average they lie
in the middle of each layer of collision cells. In the limit of a small mean-free path,
$\sqrt{\frac{k_{B}T}{m}}\delta t \ll a$, the measurement of momentum
transfer is therefore limited to a single layer of collision cells, since in the collision step momentum
transfer is confined to a single collision cell. The single-layer measurements of the
viscous stress tensor, $\sigma_{xy}^{1}$, are then averaged to provide measurements over two layers ($\sigma_{xy}^{2}$),
four layers ($\sigma_{xy}^{4}$) and six layers ($\sigma_{xy}^{6}$). We denote the corresponding
probability distributions $P^{n_{l}}(\sigma_{xy}^{n_{l}})$ and the probability functions ${\cal P}^{n_{l}}(\tau,V)$.
\par
The measurement of $\sigma_{xy}^{n_{l}}$, where $n_{l}$ is the number of measurement layers, corresponds to a measurement of entropy production in a volume $n_{l} L^{2} a/2$.
This is due to the fact that not only is the measured momentum transfer limited to a single layer of collision cells,
but within the layer the change of momentum in the two halves separated by the measurement plane differs only in sign.
We test this prediction by computing an observation volume $V(\tau)$ as
\begin{equation}
 V(\tau) = {\cal P}^{n_{l}}(\tau,V) / \tau\, .
 \label{eq:effective_volume}
\end{equation}
In the fluctuation theorem holds, then ${\cal P}^{n_{l}}(\tau,V)$
should be constant as a function of $\sigma^{n_{l}}_{xy}$.
\par
We first test these predictions for an averaging time corresponding to a single collision time, $\tau = \delta t$.
The results are shown on Fig.~\ref{fig:scaling_volume}. We find that the probability ratio computed from the probability distributions (top left) is indeed a constant (top right) and equal to the volume of measurement in
all cases (bottom). Thus, in the short-time limit, the fluctuation theorem holds.
\par
Next, we compute the running averages of $\sigma_{xy}^{n_{l}}(t)$ over different time frames $\tau \in [2\delta t; 500\delta t]$,
yielding a function $\sigma_{xy,\tau}^{n_{l}}(t)$, where $\tau$ denotes
the averaging (measurement) time. As before, we express the
measurement results in terms of the observation volume $V(\tau)$.
\par
\begin{figure*}[t!]
 \begin{center}
  \includegraphics[width=\linewidth]{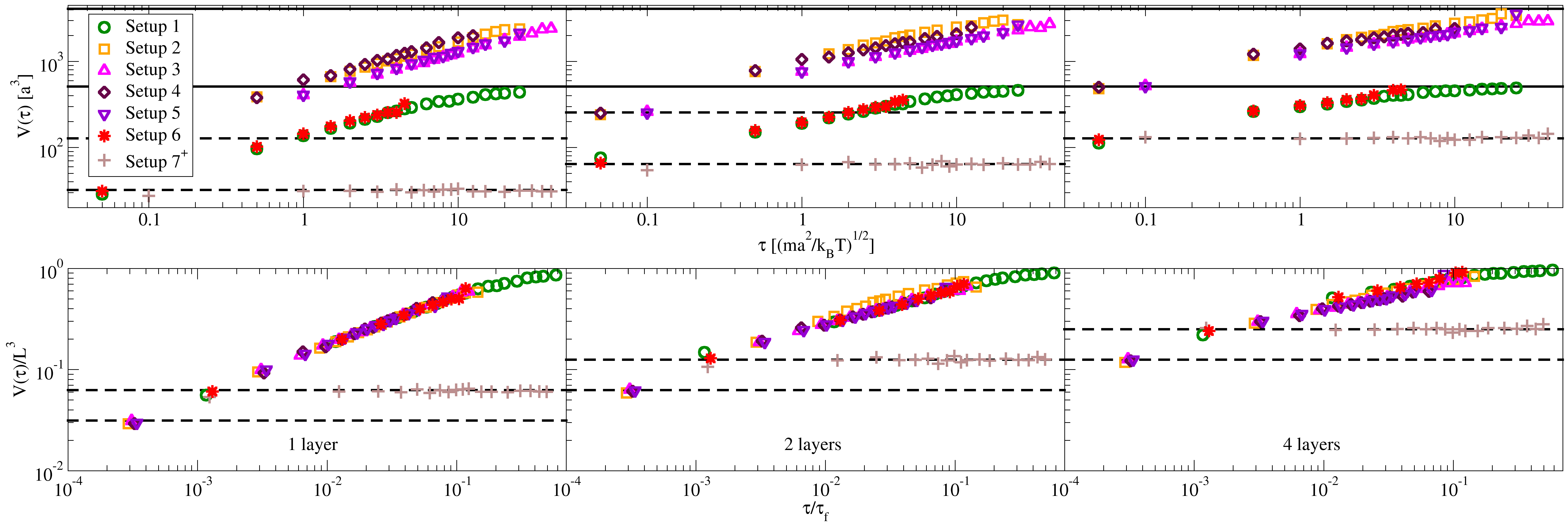}
 \end{center}
  \caption{\label{fig:scaling}(Color online) The observation volume $V(\tau)$ as a function of the averaging time $\tau$ (top)
and the normalized observation volume $V(\tau)/L^{3}$ as a function of the normalized averaging time $\tau/\tau_{f}$ (bottom)
for different setups, denoted by different symbols. The solid lines indicate the total system volumes $L^{3}$. The dashed lines indicate the single-,
two- and four-layer volumes $L^{2} n_{l} a/2$. Shown are measurements of the viscous stress tensor over one plane (left), two planes (middle) and four planes (right).}
\end{figure*}
The results for $V(\tau)$ obtained from simulations are shown on Fig.~\ref{fig:scaling} (top). For the hydrodynamic setups (1-6)
we find a surprising result. The observation volume $V(\tau)$ is not constant, at least for small values of $\tau$. This implies
that the relation~(\ref{eq:shear_fluctuation_theorem}) does not hold in this regime. The observation volume scales from the
volume within which
the measurement is performed to the total system volume $L^{3}$. We investigate this scaling in terms of hydrodynamic momentum
transport, as this is the only transport mechanism in our systems. Hydrodynamic momentum transport can be
associated with a time scale related to the propagation of momentum across a distance equal to the lateral simulation box
size $\tau_{f} = L^{2}/\nu$, where $\nu$ is the kinematic viscosity of the solvent. By scaling $V(\tau)$ by the total
volume $L^{3}$ and time by the hydrodynamic time $\tau_{f}$, we observe that, for a given number of measurement planes,
all curves collapse onto the same universal curves (Fig.~\ref{fig:scaling}, bottom). Therefore, while in the short-
and the long-time limits the fluctuation theorem is satisfied, the volume entering Eq.~(\ref{eq:shear_fluctuation_theorem}) is different. In the short-time limit it corresponds to the measurement volume, whereas at long times it corresponds to the total volume of the system. At intermediate times when $V(\tau)\ll L^{3}$, $V(\tau)$ exhibits a $\tau^{1/2}$ scaling behavior indicative of simple diffusion in the $y$-direction, $y^{2} \sim t$, and the fluctuation theorem is also satisfied.
\par
On the other hand, in the absence of transport phenomena, one expects the observation volume to be equal to the volume of measurement for all averaging times. We test this by performing simulations without hydrodynamic interactions.
To this end, we choose to adopt the following scheme. After each collision step, the velocities of the particles are randomized.
The velocity components in the off-shear-directions are sampled from a Gaussian distribution with mean zero and variance
$\sqrt{k_{B}T/m}$. The velocity component in the shear direction is sampled from a Gaussian distribution with mean
$\Big[\mathrm{d}v_{x}/\mathrm{d}y\Big](y_{i}-L/2)$ and variance $\sqrt{k_{B}T/m}$. Here $y_{i}$ denotes the $y$-component
of the coordinate of the solvent particle. This procedure guarantees a correct
shear gradient without introducing correlations within the solvent.
\par
The results for the non-hydrodynamic case, setup $7^{\dagger}$, are shown on Fig.~\ref{fig:scaling_volume} and Fig.~\ref{fig:scaling}
alongside the results for the hydrodynamic setups 1-6. Indeed, we observe that while the scaling of the observation volume with the
number of planes $n_{l}$ is the same (Fig.~\ref{fig:scaling_volume}) in all cases, the non-hydrodynamic setup does not exhibit any
scaling of the observation volume with the averaging time (Fig.~\ref{fig:scaling}), in sharp contrast to the hydrodynamic
case. This is a direct consequence of the causality principle: in the absence of transport phenomena, it should not be possible
to make conclusions about parts of the system which lie outside of the physical volume of measurement.
\par
We have shown that the fluctuation theorem~(\ref{eq:fluctuation_theorem}) holds in the presence of hydrodynamic interactions at all averaging times for solvents of different densities and viscosities undergoing shear at different shear rates. However, transport phenomena have a very significant influence on the measurement of fluctuations. In particular, we have demonstrated that the interpretation of results pertaining to measurements of entropy at short times and within small volumes in systems with hydrodynamic interactions would depend strongly on the properties of the solvent and the measurement time. Furthermore, we expect that for dynamic probes, i.e. a Brownian particle suspended in a solvent, an even finer analysis will be required, as here the observation volume will depend not only on the properties of the solvent, but also on the properties of the probe, such as its diffusion coefficient.
We expect that, since these results are a direct consequence of the causality principle, they are readily generalizable to other systems which feature correlations in combination with transport phenomena.
\begin{acknowledgements}
The authors thank Prof. Giovanni Gallavotti for encouragement and constructive criticism of the manuscript.
They also thank Prof. Roland Winkler for valuable comments and discussions.
RL would like to thank the Ecole Polytechnique F\'{e}d\'{e}rale de Lausanne for the kind hospitality, during which part of this work was performed.
MB and GF acknowledge financial support by the Swiss National Science Foundation (grant no. PP0022 119006).
GF acknowledges the support of the "Ville de Paris" for his stay in Paris during the preparation of this manuscript.
\end{acknowledgements}

\end{document}